\documentclass[fleqn,usenatbib]{mnras}
\usepackage{graphicx}
\usepackage{amsmath}
\usepackage{amssymb}
\usepackage{xspace}
\usepackage[T1]{fontenc}
\usepackage{txfonts}
\usepackage{color}
\usepackage[normalem]{ulem}

\DeclareRobustCommand{\VAN}[3]{#2}
\let\VANthebibliography\thebibliography
\def\thebibliography{\DeclareRobustCommand{\VAN}[3]{##3}\VANthebibliography}

\newcommand*{\GalSim}{\textsc{GalSim}\xspace}

\title[Sky subtraction]{Strategies for optimal sky subtraction in the low surface brightness regime}

\author[A. E. Watkins et al.]{
Aaron E. Watkins,$^{1}$\thanks{E-mail: a.watkins@herts.ac.uk (AEW)}
Sugata Kaviraj,$^{1}$
Chris C. Collins,$^{2}$
Johan H. Knapen,$^{3,4}$
Lee S. Kelvin$^{5}$,
\newauthor{
Pierre-Alain Duc$^{6}$,}
Javier Rom\'{a}n$^{7,3,4}$,
and J. Christopher Mihos$^{8}$\\
$^{1}$Centre for Astrophysics Research, University of Hertfordshire, College Lane, Hatfield AL10 9AB, UK\\
$^{2}$Astrophysics Research Institute, Liverpool John Moores University, IC2 Building, Liverpool Science Park, 146 Brownlow Hill, Liverpool L3 5RF, United\\ Kingdom\\
$^{3}$Instituto de Astrof\'{i}sica de Canarias, V\'{i}a L\'{a}ctea S/N, E-38205 La Laguna, Spain\\
$^{4}$Departamento de Astrof\'{i}sica, Universidad de La Laguna, E-38206 La Laguna, Spain\\
$^{5}$Department of Astrophysical Sciences, Princeton University, 4 Ivy Lane, Princeton, NJ 08544, USA\\
$^{6}$Universit\'{e} de Strasbourg, CNRS, Observatoire astronomique de Strasbourg, UMR 7550, F-67000 Strasbourg, France\\
$^{7}$Kapteyn Astronomical Institute, University of Groningen, Landleven 12, 9747 AD Groningen, The Netherlands\\
$^{8}$Department of Astronomy, Case Western Reserve University, 10900 Euclid Avenue, Cleveland, OH 44106, USA
}

\begin{document}
\label{firstpage}
\pagerange{\pageref{firstpage}--\pageref{lastpage}}
\maketitle

\begin{abstract}
The low surface brightness (LSB) regime ($\mu_{g} \gtrsim 26$ mag arcsec$^{-2}$) comprises a vast, mostly unexplored discovery space, from dwarf galaxies to the diffuse interstellar medium.  Accessing this regime requires precisely removing instrumental signatures and light contamination, including, most critically, night sky emission.  This is not trivial, as faint astrophysical and instrumental contamination can bias sky models at the precision needed to characterize LSB structures.  Using idealized synthetic images, we assess how this bias impacts two common LSB-oriented sky-estimation algorithms: 1.) masking and parametric modelling, and 2.) stacking and smoothing dithered exposures.  Undetected flux limits both methods by imposing a pedestal offset to all derived sky models.  Careful, deep masking of fixed sources can mitigate this, but source density always imposes a fundamental limit.  Stellar scattered light can contribute $\sim28$--$29$ mag arcsec$^{-2}$ of background flux even in low-density fields; its removal is critical prior to sky estimation.  For complex skies, image combining is an effective non-parametric approach, although it strongly depends on observing strategy and adds noise to images on the smoothing kernel scale.  Preemptive subtraction of fixed sources may be the only practical approach for robust sky estimation.  We thus tested a third algorithm, subtracting a preliminary sky-subtracted coadd from exposures to isolate sky emission.  Unfortunately, initial errors in sky estimation propagate through all subsequent sky models, making the method impractical.  For large-scale surveys like LSST, where key science goals constrain observing strategy, masking and modelling remains the optimal sky estimation approach, assuming stellar scattered light is removed first.
\end{abstract}




\section{Introduction}\label{sec:intro}

In the last few decades, large observational surveys have transformed our understanding of how the Universe evolves over cosmic time.  Those with wide areas, like the Sloan Digital Sky Survey \citep[SDSS;][]{york00, Abazajian2009}, have imaged hundreds of thousands of objects within their footprints, enabling us to quantify, in unprecedented detail, how the statistical properties of galaxies evolve over time. However, the knowledge we have gained from these surveys is naturally constrained by the objects and structures that are brighter than their surface-brightness limits\footnote{For example, the surface-brightness limit of standard-depth SDSS images is $\sim26.5$ mag arcsec$^{-2}$ \citep[$3\sigma$ within 12\arcsec \ radius apertures;][]{trujillo16}, with galaxy completeness in the SDSS catalogues decreasing rapidly for objects with effective surface brightnesses fainter than $\sim24.5$ mag arcsec$^{-2}$ \citep[e.g.,][]{Strauss2002}.}.

Indeed, most objects in the Universe are actually fainter than the surface-brightness limits of past wide-area surveys. Examples include dwarf galaxies at cosmological distances, which dominate the galaxy number density at all epochs \citep[ e.g.,][]{driver94, blanton05, baldry08, mcnaught14, Martin2019, Davis2022}, merger-induced tidal features and stellar streams, which are critical for quantifying galaxy assembly histories and testing our hierarchical paradigm \citep[e.g.,][]{Kaviraj2014, Duc2015}, intra-group and intra-cluster light, which can host much of the baryonic matter in these dense environments \citep[e.g.,][]{Burke2012, Montes2021}, and non-stellar emission or reflection features like Galactic cirrus, which illustrate the composition and cooling physics of dust grains in the interstellar medium \citep[e.g.,][]{beichman87, szomoru98, miville16, roman20, smirnov23}.

Our relative ignorance of this faint, low-surface-brightness (LSB) regime means that our empirical knowledge of the Universe is restricted to bright objects \citep[e.g., massive galaxies;][]{Jackson2021} outside the Local Universe (where LSB objects can most easily be resolved into stars). More importantly, our theoretical models are largely calibrated to these objects \citep[e.g.][]{Kaviraj2017}, making our understanding of the physics of structure formation potentially highly incomplete. Observationally studying the LSB regime, and confronting these observations with theory, is therefore critical for understanding the evolution of the Universe as a whole. 

The advent of a new era of large surveys, that are both deep and wide---for example, the Hyper Suprime-Cam Subaru Strategic Program \citep[HSC-SSP,][]{Aihara2018}, Euclid \citep{Laureijs2011}, and Rubin Observatory's Legacy Survey of Space and Time \citep[LSST;][]{Robertson2019}---promises to revolutionise our understanding of the LSB Universe. The HSC-SSP offers $\sim1400/30/4$ deg$^2$ in $grizy$ down to a $5\sigma$ point-source depth of $r\sim26/27/28$~AB mag ($\sim4/5/6$ magnitudes deeper than the SDSS), with a median seeing of 0.6 arcseconds. The Euclid Wide Survey will cover $\sim15,000$~deg$^{2}$ in visible light to a $5\sigma$ point-source depth of $m_{I,E}=26.2$~AB mag and three near-infrared bands to depths greater than 24~AB mag \citep{scaramella22}.  LSST is expected to provide, already in its commissioning phase, $\sim1000$ deg$^2$ in $(u)grizy$ to a $5\sigma$ point-source depth of $r \sim 26.5$~AB mag and $\sim$100 deg$^2$ to $r \sim 28$~AB mag ($\sim6$ magnitudes deeper than the SDSS), with a median seeing of 0.67 arcseconds.  Likewise, the full LSST Wide-Fast-Deep survey will offer $\sim$18,000~deg$^2$ to a $5\sigma$ point-source depth of $r \sim 28$~AB mag by the early 2030s. Such surveys will facilitate an unprecedented empirical exploration of the LSB Universe and produce a step change in our understanding of the evolution of the Universe.  

Notwithstanding their exceptional promise, the ability of such surveys to access their discovery space (i.e., the LSB Universe) depends critically on aspects of the pipeline data processing applied to the images. In particular, LSB objects and structures are acutely sensitive to sky over-subtraction \citep[e.g.][]{Aihara2018}. Accurate preservation of astrophysical LSB flux by the pipeline sky-subtraction algorithm is, therefore, essential for such surveys to achieve their theoretical surface-brightness limits.  In the case of LSST, for example, this will be $\sim30$ mag arcsec$^{-2}$ in the $g$, $r$, and $i$ bands, \citep[on 10\arcsec$\times$10\arcsec \ scales, about $\sim$5 mag arcsec$^{-2}$ deeper than SDSS using the same metric;][]{trujillo16}.

Night sky emission arises from a variety of sources.  Natural airglow in the visible spectrum is primarily the result of molecular or atomic oxygen lines \citep{broadfoot68, massey00}, and in the near-infrared results mainly from vibrational and rotational OH emission \citep{meinel50}.  Artificial sources contribute as well, including mercury and sodium emission lines from street lamps \citep[e.g.,][]{osterbrock92, massey00}, with intensity depending on the proximity to nearby cities.  A non-negligible amount of the sky brightness also derives from tropospheric scattering of astronomical sources, including stars, galaxies, and moonlight \citep{leinert98}, making the estimation and removal of the night sky from images non-trivial.  A complete summary of the sources of night sky emission can be found in \citet{roach73}.

Sky subtraction in past large surveys, like SDSS, was optimized primarily for object deblending, by measuring skies in very small ($128\times128$~px or $256\times256$~px) regions and interpolating or fitting smooth splines or functions across these measurements to produce sky models spanning the telescope's focal plane\footnote{\url{https://www.sdss.org/dr17/algorithms/sky/}}.  This removes much of the diffuse flux connecting bright objects, isolating more compact sources and improving small, compact object detection and photometry for catalogue creation \citep[see also][]{aihara18b}.

Much of this diffuse flux is astrophysical in origin, however, and so algorithms intent on preserving this flux require a different approach.  Typical LSB-focused studies avoid high-order fits or otherwise highly tailored sky models, to reduce the risk of over-fitting and including undetected astrophysical LSB flux in the sky models \citep[e.g.,][]{feldmeier02, rudick10, fliri16, watkins16, Montes2021}.  This is typically done in combination with aggressive masking of detected sources, often employing specialized LSB-optimized detection software to expand masks deep into the background noise \citep[e.g.,][]{akhlaghi15}.  However, as surveys begin to probe to progressively deeper limits, the density of detected sources may become so high that such masking leaves too few pixels for robust fits.  In such cases, alternative approaches become necessary, such as combining dithered images to generate non-parametric average sky models for an observing run \citep[e.g.,][]{ferrarese12, Duc2015, eigenthaler18}, or attempting to more precisely remove the influence of undetected flux prior to sky estimation, either through adaptive masking \citep[e.g.,][]{ji18} or subtraction of parametric source models from images \citep[e.g.,][]{kelvin23}. 

In ground-based near-infrared imaging, where sky is significantly brighter than the targets of observation more frequently than it is in the visible light regime, a sophisticated technique is employed in which both the telescope's primary mirror is "chopped" at a high frequency back and forth between the target field and an adjacent field, while the telescope itself is "nodded" from the target to another field a similar distance away on the opposite side.  The night sky is isolated by subtracting paired chopped exposures at each nod position, then subtracting the resulting nodded difference images from each other.  Ultimately, the goal of each method is to separate, as cleanly as possible, the time-variable night-sky emission from that of fixed astrophysical sources, but to-date, few investigations have been done to quantify the sensitivity of each proposed method to contaminating astrophysical flux.

In this paper, we assess, using idealized synthetic images, the performance of two sky-subtraction algorithms frequently utilized in LSB-oriented data reduction schemes---masking and model-fitting, and generation of averaged sky models from dithered exposures---in order to determine their viability in up-coming deep surveys such as LSST.  We also investigate the viability of a third, experimental approach meant to cleanly remove fixed sources prior to sky-estimation: the isolation of night-sky emission in individual exposures via the subtraction of a preliminary sky-subtracted image coadd.  In Section \ref{sec:synthetic}, we describe our synthetic image generation process.  Section \ref{sec:experiments} describes the four experiment cases we produce using these images.  We showcase the tests we conduct on the sky-subtraction algorithms using these images in Section \ref{sec:tests}.  We discuss the broader implications of the test results in Section \ref{sec:discussion}, and we summarize these points in Section \ref{sec:conclusions}.

\section{Synthetic images}\label{sec:synthetic}

\begin{figure*}
    \centering
    \includegraphics[scale=1.0]{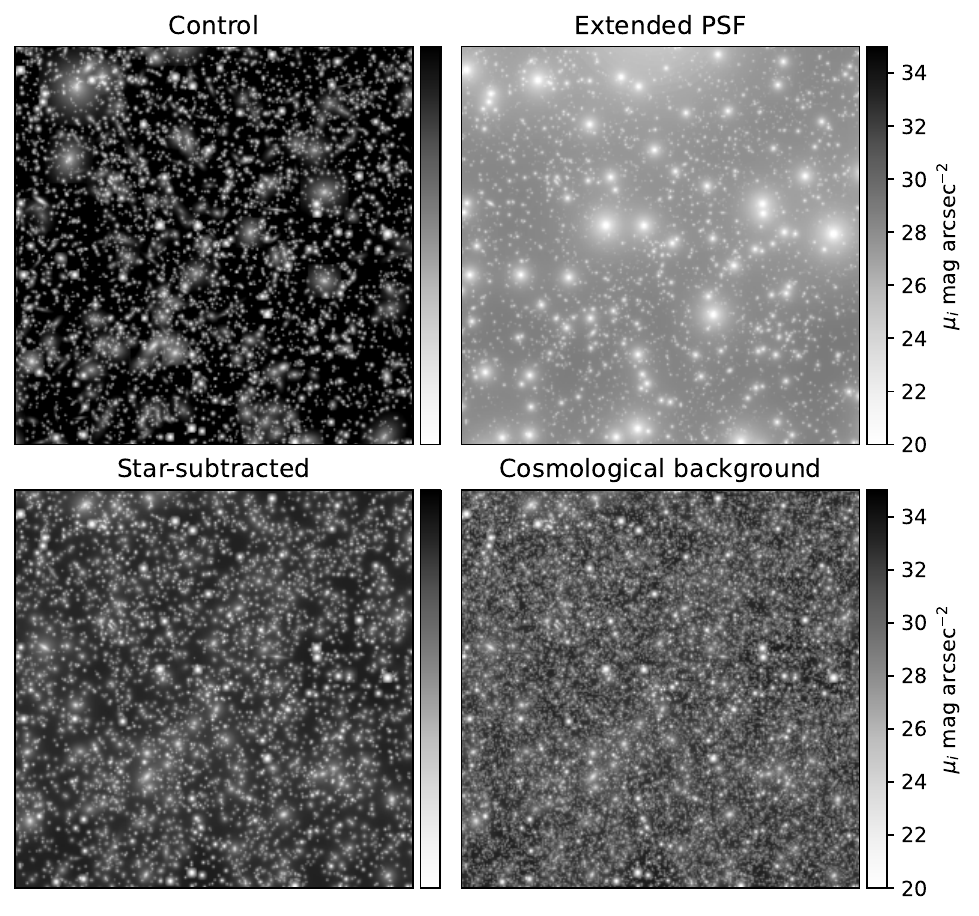}
    \caption{Example images showing four experiments in source injection.  Each image is $4096\times4096$~px ($\sim13.7 \times 13.7$~arcmin$^{2}$), and is populated with model stars and galaxies (noise-free, no model skies).  Flux is converted to units of surface brightness in the $i$-band (roughly AB magnitudes; see text).  Each panel shows a different experiment type, which are described in Sec.~\ref{sec:synthetic}. These images are idealized, such that we exclude many sources of contamination likely to influence sky estimation in real images (e.g., internal reflections, Galactic cirrus, etc.).}  \label{fig:experiments}
\end{figure*}

\begin{table}
    \centering
    \caption{Summary of six master images. $^{a}$\GalSim \ COSMOS training sample; $^{b}$Moffat parameters are $\beta=3$, FWHM=0.7\arcsec; $^{c}$12\arcmin \ HSC PSF from \citet{Montes2021}; $^{d}$positions and magnitudes from Data Release 12 \citep{alam15} and Type$=1$ sources from \citet{laigle16}; $^{e}$\citet{dubois14}}
    \label{tab:masters}
    \begin{tabular}{lll}
        \hline\hline
        Master Image & Source Type & PSF \\
        \hline \\
        1 & galaxies$^{a}$ & Moffat$^{b}$ \\
        2 & galaxies$^{a}$ & M21$^{c}$ \\
        3 & stars$^{d}$ & Moffat$^{b}$ \\
        4 & stars$^{d}$ & M21$^{c}$ \\
        5 & Hz-AGN galaxies$^{e}$ ($i>25.2$) & Moffat$^{b}$ \\
        6 & Hz-AGN galaxies$^{e}$ ($i>26.1$) & Moffat$^{b}$ \\
    \end{tabular}
\end{table}

To properly quantify the performance of different sky-subtraction algorithms, we require full control over the kinds of sources and sky backgrounds present in our images.  We thus create sets of fully synthetic images, populated with model galaxies, stars, and skies.  To generate fake observations, we first create six master images spanning $10240\times 10240$~px, each with a world coordinate system centred at $\alpha=150.040635^{\circ}$ and $\delta=2.208592^{\circ}$, near the centre of the Hubble Space Telescope (HST) Cosmic Evolution Survey (COSMOS) footprint, and with a 0.2 arcsec px$^{-1}$ pixel scale to mimic that of LSSTCam \citep{roodman18}.  Fake observations are then drawn from different combinations of these master images, to which we add model noise and skies.

To simulate a variety of experimental conditions (Sec.~\ref{sec:experiments}), we populate these master images with different kinds of objects using the open-source model-generation software package \GalSim \ \citep{rowe15}.  In the first, simulating a normal population of extended sources, we inject only mock galaxies, drawing each as a single S\'{e}rsic profile ($n=0.3$--$6.2$, limited by the \GalSim \ software) convolved with a Moffat profile \citep{moffat69} kernel, with $\beta=3$ and FWHM$=0.7$\arcsec \ to simulate LSST-like seeing.  We obtain mock S\'{e}rsic parameters, celestial coordinates, and magnitudes from the HST-COSMOS training sample available through \GalSim\footnote{\url{https://galsim-developers.github.io/GalSim/_build/html/real_gal.html}}.  This source catalogue is limited to $i$-band magnitudes $<25.2$, hence it excludes many high-redshift objects.  We draw all mock galaxies to their projected $\mu_{i}=35$ mag arcsec$^{-2}$ isophotes, to ensure a realistic contribution of galactic LSB flux.  To fill in holes in the catalogue data induced by bright star masks, we randomly perturb the catalogue source coordinates by adding offsets drawn from a uniform distribution with a range of $\pm 30$\arcsec to the right ascension and declination values.

Scattered light from extended sources being a possible contaminant for sky models, our second master image contains the same galaxy profiles, but convolved instead with the Subaru Telescope Hyper Suprime-Cam \citep[HSC;][]{nakaya08} extended PSF measured to 7\arcmin \ radius by \citet[hereafter, M21]{Montes2021}.  We use that study's $g-$band PSF for our work, as its core is less noisy than that of the $i-$band PSF, though we find our broad results to be insensitive to this choice.

To simulate Milky Way sources, we inject only mock stars into the third master image, using the same Moffat profile we used to convolve our mock galaxies in our first master image.  We obtain celestial coordinates and magnitudes for our mock stars from the SDSS Data Release 12 \citep{alam15} to sample bright stars (the brightest being $m_{i} = 8.17$), and using sources from the COSMOS2015 catalogue \citep{laigle16} to sample fainter stars.  From the latter, we select only sources with parameter Type $=1$ (indicating a star-like profile), and we exclude those with coordinates within 2\arcsec of stars in the SDSS catalogue to avoid redundancies.  we fix each star's stamp size to $100\times100$~px to avoid noticeable cutoffs at the stamp boundaries of the brightest stars.  Ringing artifacts generated around bright stars using fast Fourier transform convolution create background noise which noticeably impacts our results; hence, we convolve our sources in real space rather than Fourier space.  Such noise around faint sources (including our mock galaxies) is negligible.

Mirroring the purpose of the second master image, but for stellar scattered light, our fourth master image contains the same mock star sources described for the third master image, but injected instead using the M21 PSF.  We likewise inject these stars in real space, using a PyRAF \citep[Version 2.1.15;][]{pyraf} script, which scales the M21 PSF image in flux using the IMARITH task and places the scaled star model at the proper coordinates in the master image using the IMSHIFT task (with a linear interpolation for partial pixel shifts).

Source crowding is also a potential problem for sky estimation in deep imaging surveys; hence, we inject mock high-redshift galaxies into our fifth master image, convolved with the Moffat profile described above.  Celestial coordinates and $i$-band magnitudes for these sources come from a Horizon-AGN simulation \citep[hereafter, Hz-AGN;][]{dubois14} light cone \citep{gouin19, laigle19}, which we offset in coordinates to the centre of our mock field of view.  This catalogue contains photometric information for all simulated galaxies with stellar mass $\mathcal{M}_{*} > 10^{9} \mathcal{M}_{\odot}$ out to $z=4$, thus simulating high-redshift sources likely detectable in a reasonably deep cosmological survey.  We include only Hz-AGN sources with $i$-band magnitudes $>25.2$, to fill in the gaps of the COSMOS training sample.  To save processing time, as this sample contains $>450,000$ objects, we inject these galaxies as point sources in the same manner as the stars rather than as S\'{e}rsic profiles.  For objects this faint, the difference between Moffat profiles and the M21 profile is also negligible.

Finally, to assess the influence of undetected faint sources, we create a sixth master image identical to the fifth, but including only Hz-AGN sources with $i-$band magnitudes $>26.1$.  We use this only to isolate the brighter Hz-AGN sources in our experiments, by subtracting this image from the fifth prior to generating object masks, for example.

\begin{figure}
    \centering
    \includegraphics[scale=1.0]{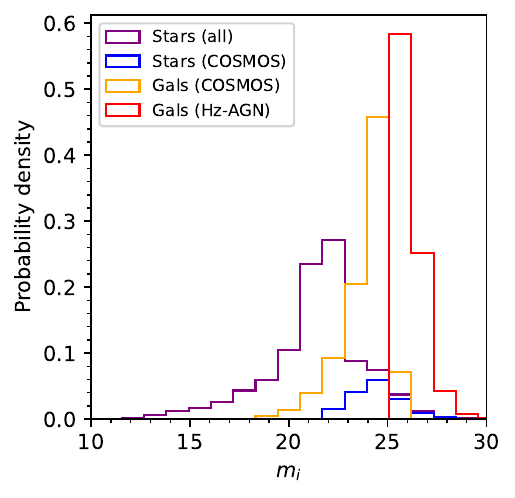}
    \caption{Histograms of synthetic source magnitudes, delineated by source type.  All histograms are scaled by their respective full sample sizes, save that for COSMOS stars, which is scaled by the full star sample size. \label{fig:sources}}
\end{figure}

Table~\ref{tab:masters} summarizes our six master images.  We show the distribution of magnitudes for all synthetic sources in Fig.~\ref{fig:sources}.  In total, our combined COSMOS galaxy, stellar, and Hz-AGN catalogue of synthetic sources contains 896250 objects within a 1~deg$^{2}$ area, of which 54923 are COSMOS galaxies (with source density per unit area, $N/A$, peaking for objects with $m_{i} \approx 24$), 51973 are stars (with $N/A$ peaking at $m_{i} \approx 21$), and the remaining 789354 are Hz-AGN galaxies (with $N/A$ peaking at $m_{i} \approx 25$).  As each master image is only $\sim 0.32$~deg$^{2}$, not all of these catalogue objects are present in each master image, but the source densities for each source type subset are similar between the master images and the full catalogue.

\section{Experiments}\label{sec:experiments}

\begin{table}
\centering
\caption{Summary of synthetic sources and purposes for four experiment types.  Master Image IDs are from Table~\ref{tab:masters}.}
\label{tab:experiments}
\begin{tabular}{lll}
\hline\hline
Experiment & Masters & Purpose \\
\hline \\
Control & 1,3 & Idealized baseline \\
Extended PSF &  2,4 & Scattered light contamination \\
Star-subtracted & 2,3 & Simulating LSB-processing \\
Cosmological background & 2,3,5 & Undetected sources, crowding \\
\hline
\end{tabular}
\end{table}

By adding together combinations of the master images described in the previous section, we conduct four experiments.  The first is a baseline, optimal scenario, using only mock HST-COSMOS galaxies (hereafter, COSMOS galaxies) and mock stars, both convolved with our chosen Moffat profile.  The second scenario investigates the impact of the wings of the PSF by combining the master images containing COSMOS galaxies and stars convolved with the M21 PSF.  Our third experiment includes the M21 PSF--convolved COSMOS galaxies and the Moffat profile stars, mimicking the expectation for an LSB-optimized data-reduction scheme that would have subtracted the halos of bright stars \citep[e.g.,][with the caveat that real star-subtraction algorithms do not remove bright stars nearly as cleanly]{slater09, infantesainz20}.  Finally, to investigate the impact of the cosmological background \citep[effectively, extragalactic background light, or EBL; e.g.,][]{partridge68, hara74, madau00, bernstein07, driver16} on an LSB-processed image, we combine images with COSMOS galaxies convolved with the M21 PSF, all Hz-AGN galaxies, and Moffat profile stars.  We refer to each of these experiments, respectively, as control, extended PSF, star-subtracted, and cosmological background.  We show example images for each experiment type in Fig.~\ref{fig:experiments}, and we summarize the experiment parameters in Table~\ref{tab:experiments}, including the master images combined to create each experiment case (referring to Table~\ref{tab:masters}).

These combined masters serve as synthetic night skies, from which we conduct mock observations by sampling smaller images.  Each such mock exposure has dimensions of $4096\times4096$px ($\sim13.7 \times 13.7$~arcmin$^{2}$), always centred at pixel coordinates within the master images at least $2048$~px from each edge.  To simulate more realistic exposures, we add mock instrumental noise to each image using \GalSim's CCDNoise module, using a gain of 3 $e^{-}$ ADU$^{-1}$ and a read noise of 4.5 $e^{-}$, based on the properties of an average CCD in HSC\footnote{\url{https://www.subarutelescope.org/Observing/Instruments/HSC/parameters.html}}.  We add either a constant pedestal value to each image, or a polynomial model generated using Astropy's \citep{astropy:2018} Legendre2D modelling software, to emulate night sky emission, including Poisson noise in both varieties (as arises in the case of real integrated sky flux).

We work throughout in analog-to-digital units (ADU), to ensure that noise characteristics are realistic with respect to injected sources.  Converting from AB magnitudes \citep{oke74} to ADU is not straight-forward for any particular survey, so we derive it for our images in the following way: using the Rubin Science Platform \citep{juric17} Butler, we obtain instrumental photometric calibrations from $i$-band HSC images of 2700 CCDs (calexp data type) drawn from a handful of visits in Tract 9813 (within the COSMOS footprint) using the photoCalib.instFluxToMagnitude function, which provides the conversion between instrumental flux and AB magnitudes for visit-level source detections.  We take the average calibration zeropoint of these exposures, $ZP=33.1$, as our calibration zeropoint.  Among all $2700$ exposures, the standard deviation among zeropoints is only $0.07$~mag, which we adopt as the characteristic uncertainty in our tests when converting from ADU to surface brightness units.

\section{Sky-subtraction tests}\label{sec:tests}
 
 This section discusses the efficacy and pitfalls of different sky-subtraction strategies under each of the four experimental cases we introduced in Sec.~\ref{sec:experiments}.  We first discuss the accuracy with which we can recover model skies from masked images using either the mean and median (for flat model skies, Sec.~\ref{subsubsec:meanmed}), or using polynomial fitting (for polynomial model skies, Sec.~\ref{subsubsec:polyfit}).  We follow this by discussing the averaged-sky technique described by \citet{Duc2015} \citep[a technique derived from earlier image-stacking strategies; e.g.,][]{tyson90}, assuming a perfectly static sky across all synthetic exposures (Sec.~\ref{subsec:avskies}).  Finally, we explore a more experimental technique, in which we attempt to remove static astrophysical sources prior to sky-estimation by first subtracting from each image a preliminary sky-subtracted image coadd (Sec.~\ref{subsec:coaddsub}).

 These simulated images are highly idealized.  They do not include many sources of contamination which are found in real astronomical images, either instrumental (reflections, saturation bleeds, etc.) or astrophysical (Galactic cirrus, galactic stellar halos, intra-cluster light, etc.).  As such, the results of these tests should be considered only as a starting point for real data reduction pipelines which outline the broad response of different sky-subtraction algorithms to different forms of flux contamination.  We discuss this in more detail in Sec.~\ref{subsec:other}.

\subsection{Masking and modelling}\label{subsec:maskmodel}

\subsubsection{Clipped mean and median}\label{subsubsec:meanmed}

\begin{figure*}
    \centering
    \includegraphics[scale=1.0]{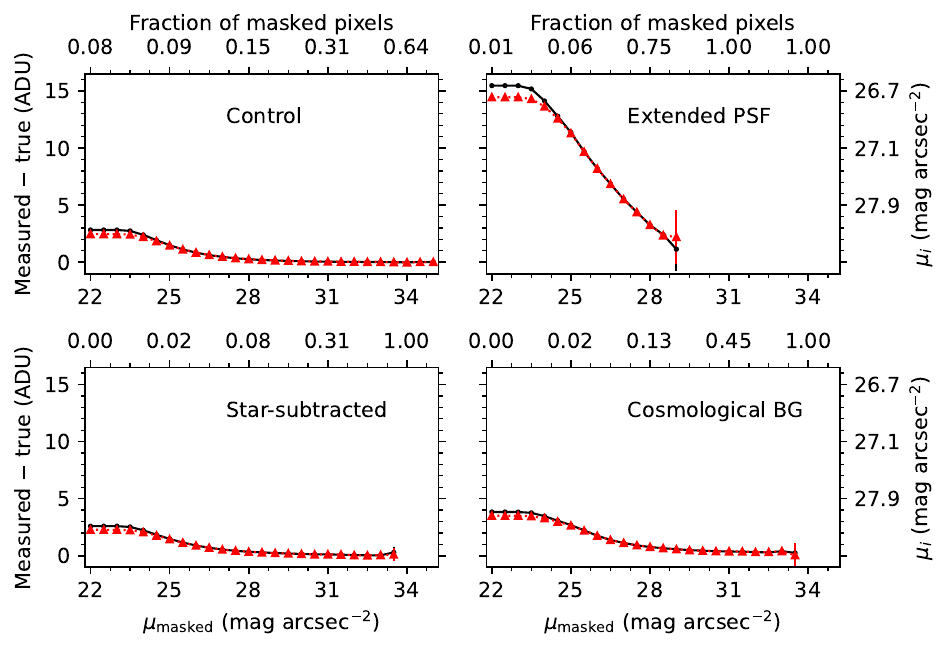}
    \caption{Testing the recovery of the mean and median values of a flat sky model with different amounts of source-masking.  In each panel, we show the results from our four experiment types (Sec.~\ref{sec:synthetic}), with solid black lines and circles showing results for a clipped mean, and with dotted red lines and triangles for a clipped median.  The x-axes show the depth of source masks (generated from the noise-free images shown in Fig.~\ref{fig:experiments}) in units of surface brightness.  Along the top, we show the fraction of masked pixels in the images when masked to each surface brightness depth.  The y-axes show the differences between the measured and the ground-truth mean and median values, measured using the same clipped mean and median algorithm using a sourceless sky image.  We convert these values into equivalent $i-$band surface brightnesses on the right using our photometric zeropoint $ZP=33.1$ and pixel scale 0.2 arcsec px$^{-1}$.  Where profiles are truncated or highly skewed, the image had no or very few pixels left with which to estimate a mean or median. We adopt as our uncertainty the standard errors on the mean or median of the masked images, as the ground-truth values are known exactly.  Most errorbars are smaller than the point size.} \label{fig:meanmedian}
\end{figure*}

Here, we investigate the robustness of the clipped mean and median to the depth to which sources are masked.  For images taken with exposure times significantly shorter than the variability time scale of the night sky in the image's photometric band \citep[5--40 minutes in the $i-$band;][]{moreels08}, night sky emission can appear flat within single exposures.  Typically, this occurs when the field of view is small and when there is no contamination from moonlight or other off-axis flux.

\citet{trujillo16} and \citet{trujillo21} show two examples of this (although it should be noted that some of the sky gradients present in images used in those two studies may have been removed via flat-fielding).  In such cases, the sky can be estimated using a single number.  In practice, this is typically the mean or median intensities of pixels deemed appropriately void of astrophysical flux.  When estimated using a clipping algorithm to reject values in the tails of the background's noise distribution, these estimators are thought to be quite robust, assuming there is no strong flux contamination from, for example, leakage through static source masks shifting the mean or median to a higher value.

We assessed the clipped mean and median as sky estimators using the synthetic exposures created as described in Sec.~\ref{sec:synthetic}.  For each experimental case, we centred the exposures at the same coordinates, injecting into each the same flat synthetic sky, with a mean flux of $6951.2$~ADU \citep[corresponding to $\mu_{i}=20$ mag arcsec$^{-2}$, roughly the $i-$band dark sky brightness expected at Cerro Pach\'{o}n;][]{yoachim16}.  We then iteratively masked all synthetic sources to an array of surface brightness depths, from $\mu_{{\rm mask},\, i}=22$ to $35$ in steps of 0.5 mag arcsec$^{-2}$.  At each mask depth, we re-estimated the skies using both the clipped mean and the clipped median, with five iterations of $3\sigma$ clipping.  For the ground-truth comparison, we measured the clipped means and medians of sourceless, sky-only images with identical noise.  While we might have compared against the true input mean value, we wished to ensure a fair comparison between measured values in the presence of variable noise.  We adopted the standard errors on the mean and median of the masked image backgrounds as our measurement uncertainty for each mask depth, taking the comparison mean and median of the reference sky as a precisely known quantity.

Fig.~\ref{fig:meanmedian} shows these comparisons for each of the four experiment types outlined in Sec.~\ref{sec:experiments}.  In each panel, we show the difference between the measured and ground-truth clipped means (medians) as the black solid (red dotted) lines with black points (red triangles) as a function of the depth to which we masked the synthetic sources in each experiment case.  The top axis of each panel shows the fraction of masked pixels in each image at each mask depth given on the bottom axis (with a value of 1.00 indicating no unmasked pixels, resulting in no estimates of the mean or median).

In each panel, we see similar behaviour.  As the depth to which synthetic sources are masked increases, the measured clipped means and medians approach the ground-truth values monotonically.  At shallow mask depth, the clipped medians (red curves) are slightly more robust to the unmasked flux than the clipped means, although at greater mask depth both curves converge.  In the control case, where LSB flux from scattered light and background objects are not included, these reach within $\sim0.01$ ADU ($\mu_{i} \approx 34.6$ mag arcsec$^{-2}$) of their respective ground-truth values by $\mu_{{\rm mask},\, i} = 33$ mag arcsec$^{-2}$.  \citet{borlaff19} conducted a similar experiment, comparing estimated sky brightnesses when masking simulated images using different detection software, and found qualitatively similar behavior (see their Fig.~9).

In other cases, when scattered light and the cosmological background are included, we see the limits of the masking approach.  For example, when we convolve both stars and galaxies with the 8\arcmin \ M21 PSF\footnote{This is the extent using our pixel scale of 0.2\arcsec \ px$^{-1}$; at the original HSC pixel scale it is 7\arcmin.} (top-right panel), the amount of scattered light limits the mask depths we can test to $\sim29$~mag arcsec$^{-2}$, beyond which depth the image is entirely masked.  At this depth, the measured clipped means and medians are still $\sim1.5$~ADU ($\mu_{i} \sim 29.2$ mag arcsec$^{-2}$) too high.  Internal reflections, not included in our models, would exacerbate this.

This contribution comes primarily from the stars, however: in the bottom-left panel, where we convolved only galaxies with the M21 PSF, the curves are able to converge to the true values by approximately the same mask depth as in the control case.  Only when the number of unmasked pixels approaches zero do these estimators lose their robustness, which in this case does not occur until $\mu_{{\rm mask},\, i}=33.5$.  The stark difference between the top-right and bottom-left panels is a clear demonstration of the importance of subtracting stellar scattered light from images as part of any LSB-oriented data reduction pipeline.  This is evident from visual inspection of the synthetic images as well; in the top-right panel of Fig.~\ref{fig:experiments}, it is obvious that scattered light from our model stars generates a complex background with a surface brightness of $\mu_{i} \approx 28$--$29$ mag arcsec$^{-2}$ even for the relatively low-density COSMOS field on which our synthetic images are based.

When investigating the impact of the cosmological background, we took a slightly different approach to masking.  Rather than masking all model sources, we excluded any galaxy sources with $m_{i} > 26.1$~mag from the masks by using our sixth master image (containing only those sources), to simulate a typical source detection algorithm's lack of ability to robustly detect faint objects \citep[this magnitude corresponds to the $5\sigma$ limit in the $i$-band in COSMOS2015;][]{laigle16}.  Consistent with the experiment conducted by \citet{ji18}, leaving these sources unmasked results in a flattening of the two curves at a value of $\sim 0.35$~ADU, equivalent to $\mu_{i} \approx 30.7$ mag arcsec$^{-2}$.  Evidently, this value is sensitive to one's particular survey detection limits, but any unmasked cosmological background will impose such a pedestal on the sky estimates even if detected sources are masked to extremely low surface brightnesses.  Thankfully, the EBL is derivable from galaxy counts \citep[e.g.,][]{metcalfe01}, hence we reiterate here \citep[as][]{ji18} the importance of subtracting an EBL pedestal from any measured sky model.

\begin{figure}
    \centering
    \includegraphics[scale=1.0]{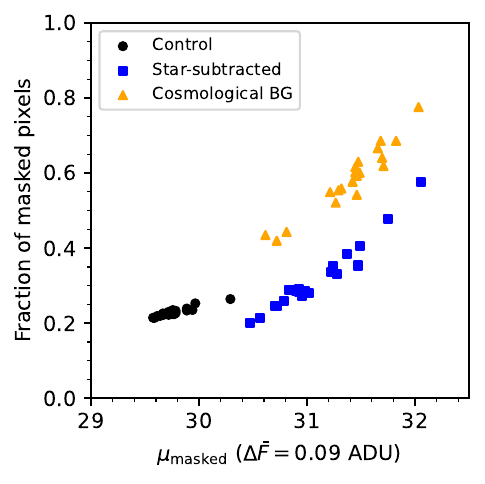}
    \caption{Depth to which mock sources were masked vs. fraction of masked pixels in mock dithered images at which the difference between the measured and ground-truth clipped mean sky background flux reached 0.09~ADU, or $\sim$10\% of the expected 10-year surface brightness depth in the $i$-band for LSST.  Each point type indicates a different experiment type (Sec.~\ref{sec:synthetic}).  The extended PSF case is missing, as in no images did the measured clipped mean converge to within 0.09~ADU of the ground-truth clipped mean.  \label{fig:convergence}}
\end{figure}

Masking sources can thus be effective for approaching the true sky value, assuming the majority of sources present in the images are masked to sufficient depth and assuming appropriate empirical corrections are made for scattered light and EBL.  This behaviour has diminishing returns, however, as the robustness of the clipped mean or median begins to falter with too few available unmasked pixels.  Fig.~\ref{fig:convergence} better illustrates this.  Here, for each experiment type, we generated 20 synthetic exposures at random coordinates within our master images, added synthetic model skies to each one, and derived the mask depths (masking all sources in all experiment types, including the cosmological background) required for the measured clipped means to converge to within 0.09~ADU ($\mu_{i}=32.2$ mag arcsec$^{-2}$) of each exposure's ground-truth clipped mean, roughly 10\% of the expected 10-year surface brightness depth in the $i$-band for LSST\footnote{\url{https://smtn-016.lsst.io/}}.  Each marker type shows a different experiment type; the extended PSF case is not present, as the clipped means never converged to within $0.09$~ADU of the ground-truth value in these images.

Generally, the larger the fraction of the image that is masked, the deeper the model sources must be masked to reach our chosen convergence limit, no matter the experiment type.  Each experiment shows a different trend.  This is due to the distribution of LSB flux in the images---because the density of LSB sources is higher in the cosmological background experiment, for example, masking all sources to a given depth masks more total pixels than the star-subtracted experiment, which for fixed image size constitutes a larger fraction of the image.  Therefore, all three experiments demonstrate that when the source density is high enough (i.e., when the image is confusion-limited with respect to the sky measurement), it becomes impossible to mask an image to deep enough levels to perfectly estimate the sky brightness even using a clipped mean (or median).

\subsubsection{Polynomial-fitting}\label{subsubsec:polyfit}

\begin{figure*}
    \centering
    \includegraphics[scale=1.0]{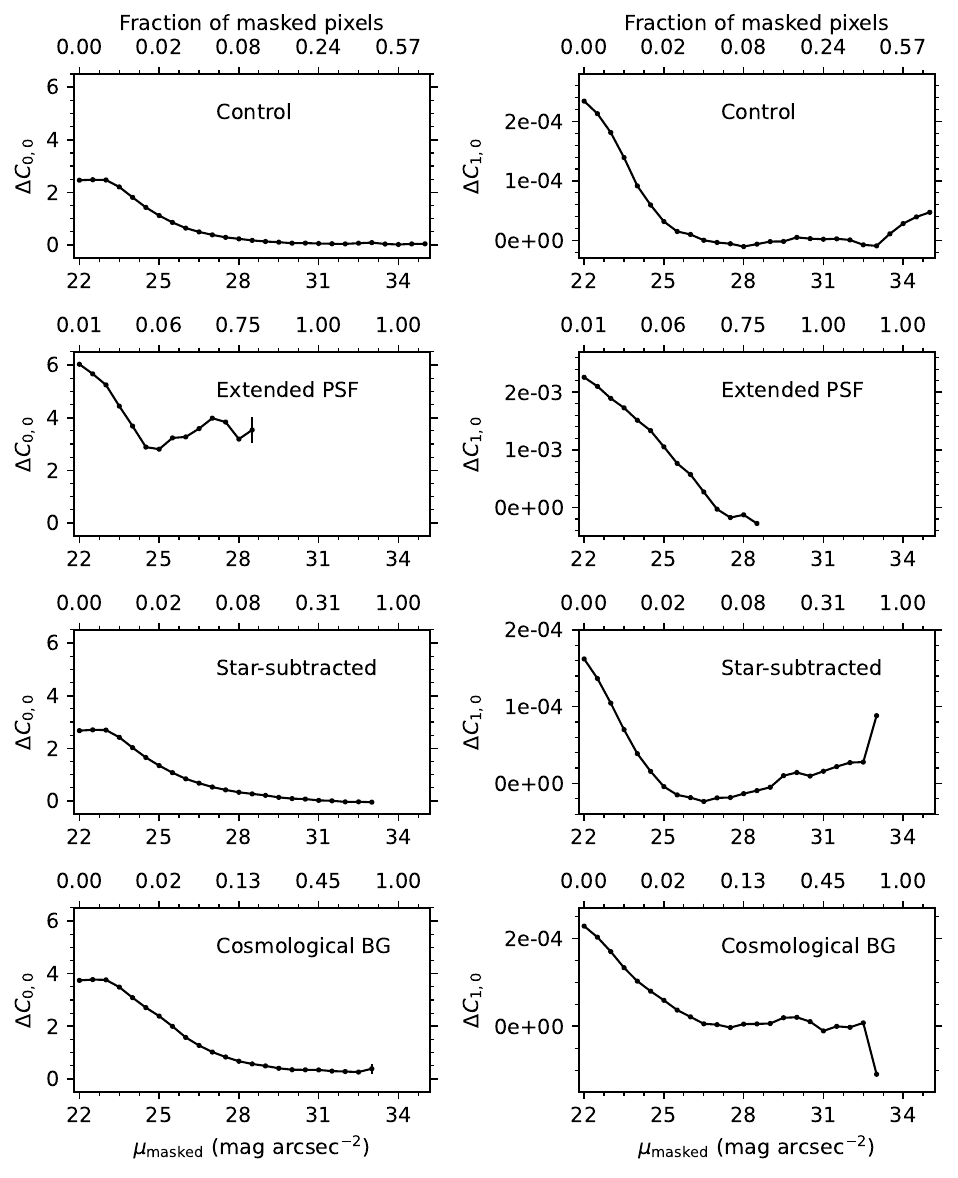}
    \caption{Comparing terms in best-fit plane (first-order Legendre polynomial) sky models between masked images and ground-truth fits.  The panels are ordered by experiment type in the manner of Fig.~\ref{fig:meanmedian}, and, similarly, curves truncate when no pixels are available for the fit.  The left panels show recovery of the moment zero term in the best-fit Legendre polynomial, while right panels show recovery of the first-order x-axis coefficients.  Y-axis limits in the right panels vary to better showcase the true shapes of the curves. Error bars on both coefficients are derived from the covariance matrix of the best-fit first-order polynomial models for each masked image.  Most are smaller than the point size (all, for the $C_{1,0}$ terms).}
 \label{fig:planefit}
\end{figure*}

We also tested the sensitivity of more complex sky-fitting procedures to source masking by injecting into each experimental case image a first-order Legendre polynomial sky model.  We built this model using Astropy's Legendre2D function, using the following coefficients (following the notation used by Astropy): $C_{0,0} = 6951.20$, $C_{1,0}=0.015$, $C_{0,1}=0.0056$, and $C_{1,1}=0.0$, a $\sim1$\% gradient across the image.  While the moment-zero term mimics that used in the mean and median test described above ($\mu_{i} = 20$ mag arcsec$^{-2}$), we chose the following terms to reflect typical best-fit sky models generated during the data reduction procedure described in \citet{watkins16} for $V-$band Burrell Schmidt Telescope imaging (15 minute exposures with a 0.9m mirror over a 1~degree$^{2}$ field of view in dark sky conditions).  The choice of values is arbitrary, but anchoring them in real data helps keep the plane amplitudes grounded.

With this mock sky injected, we mimicked the procedure described in the previous section, masking each image to an array of depths and re-estimating the sky coefficients at each step.  We fit the skies by median-binning the masked images into $64\times64$~px bins, re-fit the unmasked pixels as first-order Legendre polynomials, weighting by the fraction of unmasked pixels in each bin, then recorded each best-fit term at each step.  Mirroring how we compared the means and medians, we measured our ground-truth coefficients by performing the same polynomial fitting steps on sourceless, sky-only images with identical noise.  We show the results of this in Fig.~\ref{fig:planefit}, for the $C_{0,0}$ and $C_{1,0}$ terms in the panels in the left and right column, respectively.  Here we derived the uncertainties on each coefficient using the covariance matrices of the best-fit polynomials on the masked images.

We see similar behaviour here to that of the clipped means and medians, albeit with more variability.  The $C_{0,0}$ term being similar to the mean, we see the same gradual convergence to the true values with increasing mask depth, with profiles truncating when the fraction of masked pixels reaches 100\%.  The $C_{1,0}$ terms typically approach the ground-truth values by $\mu_{\rm masked}=27$ mag arcsec$^{-2}$, though they often diverge again beyond this depth.  However, this coefficient's measured values are rarely far from the ground-truth values.  Even in the worst-performing case, the extended PSF case with $\mu_{{\rm masked}}=22$ mag arcsec$^{-2}$, it is too high by only $\sim10$\%; in other cases, this error falls to $<1$\%.  Other cross-terms show similar behaviour, suggesting that these terms are fairly insensitive to masking, and mostly change to compensate for the changes in $C_{0,0}$.  The direction of the plane gradient thus seems robust to mask depth, for low-order polynomials, in these images.  However, Galactic cirrus or other large-scale LSB contamination not included here would likely bias the measured plane gradients and directions as well.

In the control case, $C_{0,0}$ never converges to within $0.01$~ADU of the ground-truth value; the curve instead flattens closer to $0.05$~ADU ($\mu_{i}\sim32.9$ mag arcsec$^{-2}$) by $\mu_{\rm mask}\sim31$ mag arcsec$^{-2}$, with some small scatter about this value until $\mu_{\rm mask}\sim35$, though for most surveys this level of precision should be acceptable.  Scattered light from stars, again, shows the most important impact, with polynomial fits never converging to the correct solution and showing the largest variability among all experiment cases.  The star-subtracted case reaches $0.01$~ADU of the ground-truth value by $\mu_{{\rm masked}}=31.5$ mag arcsec$^{-2}$, drifting slightly below this ($-0.05$~ADU) until the entire image is masked to 33 mag arcsec$^{-2}$ depth, while the cosmological background again imposes a pedestal of $\sim0.35$~ADU, which is reflected in the $C_{0,0}$ term.

For relatively simple sky models, polynomial-fitting shows the same general sensitivity to masking as the clipped mean and median, although the added model complexity yields more diverse behaviour.  In optical bands, flat or planar skies are good approximations of the true sky emission, even for long exposure times on degree scales \citep[e.g.,][]{watkins16}.  In redder bands, however, sky emission can show much more complex structure over much shorter timescales \citep{moreels08}.  We therefore also investigated the behaviour of higher-order polynomials, in an effort to determine what level of complexity in a sky model is acceptable under what circumstances.

\begin{figure*}
    \centering
    \includegraphics[scale=1.0]{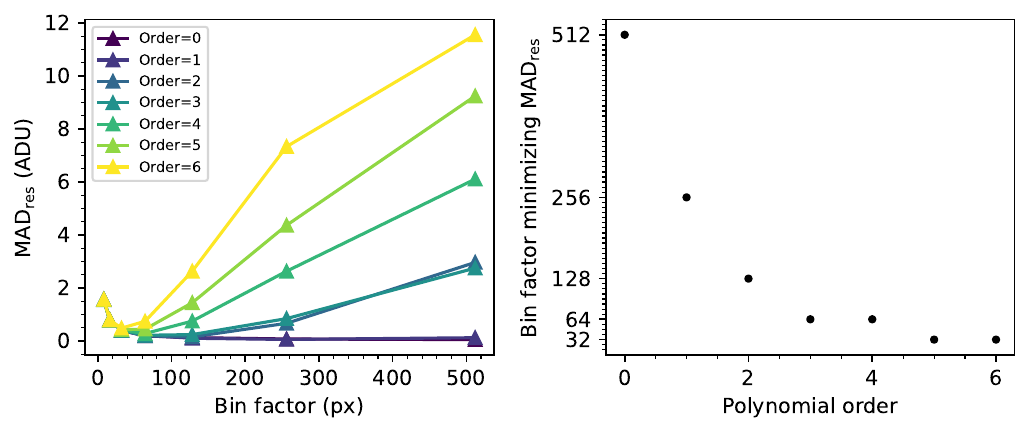}
    \caption{Showing how binning images affects recovery of sky models of varying polynomial order.  The left panel shows the median absolute deviation of the residuals of binned, noisy sky models and binned, noiseless sky models against the factor by which each image was median-binned, with each color representing a different polynomial order sky model.  The right panel shows the bin factor which minimizes the MAD for each polynomial order.  Too high a bin factor for complex skies results in poor sky models and substantial residuals.  \label{fig:polyres}}
\end{figure*}

When fitting our polynomial models, we always binned the masked images to speed up the fitting process.  This also enhances the signal to noise of each ``superpixel'' used in the fitting, making for more accurate fits.  However, for skies with complex shapes, using overly large bins risks washing out some of the skies' finer structure.  Fig.~\ref{fig:polyres} shows this in practice.  To produce this figure, we created a series of sky models with polynomial orders from $0$ to $6$ by fitting polynomials with those orders to images of pure Poisson noise.  We then created noisy versions of each model and median-binned the noiseless and noisy sky models into a range of bins with powers of $2$, from $8$~px to $512$~px.  As a metric assessing the impact of the binning on the sky models, we subtracted each noiseless sky model from each binned, noisy sky model and measured the median absolute deviation (MAD, a non-parametric measure of the dispersion) of these residual images.

We plot these values of MAD in the left panel of Fig.~\ref{fig:polyres} as a function of bin factor.  We expect the MAD to decrease with increasing bin factor as $\sim N^{-1/2}$.  However, this is only the case for the flat sky model (order $0$); for every other model, the MAD decreases to a minimum value, beyond which it increases due to large-scale differences in structure between the best-fit model and the true sky, a kind of correlated noise.

We plot the bin factor which minimizes each curve as a function of the associated sky models' polynomial orders in the right panel of Fig.~\ref{fig:polyres}.  We recreated this figure a number of different times throughout our experimentation, with similar results each time despite pseudo-randomly generating our polynomial models. This suggests that the shape of the curve is driven primarily by the relative sizes of typical features in each model with respect to the full image size.  Therefore, the right panel of Fig.~\ref{fig:polyres} could be used as a reference for the appropriate factor by which to bin images when fitting sky backgrounds as polynomials of different orders, scaled to one's specific image dimensions.  For example, if one chooses to fit one's background as a second-order polynomial, the most effective bin size to use is about $1/32$ the image axis lengths.

The unfortunate consequence of this behaviour is that the most complex skies require the smallest bin factors, leading to the noisiest sky models.  In such cases, non-parametric estimations of skies may be more useful.  We discuss one such method in the following section.

\subsection{Averaging dithered exposures}\label{subsec:avskies}

\begin{figure*}
    \centering
    \includegraphics[scale=1.0]{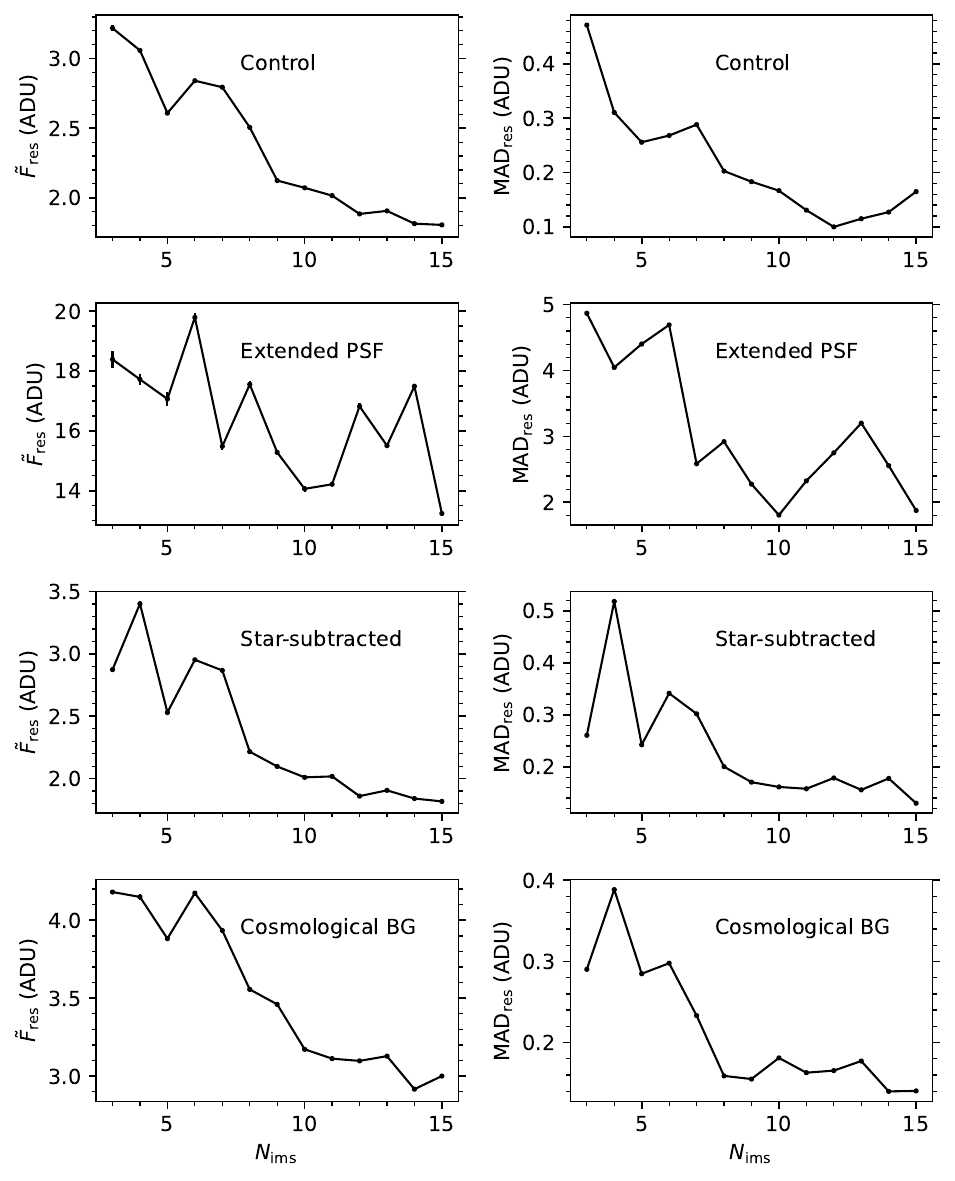}
    \caption{The median (left column) and median absolute deviation (right column) of the residuals from model sky images produced by median-combining dithered synthetic exposures, each with the same model sky, as a function of the number of exposures combined to create the sky models. Error bars, derived from the standard error on the median, are always smaller than the point size.} \label{fig:dither}
\end{figure*}

Non-parametric sky estimation provides an alternative to masking and modelling.  Through our experimentation, we found that the method employed by the Elixer-LSB pipeline \citep{ferrarese12, Duc2015}---median-combining dithered exposures and smoothing the resulting combined image with a large kernel---is an effective and fast non-parametric alternative.  This specific implementation derives from earlier work designed for LSB photometry, employing a similar shift-and-stack strategy to derive sky models post instrumental signature removal \citep[e.g.,][]{tyson88, tyson90}.  Contaminating flux also affects this approach, however.  Sources in the images combined to create the sky models tend to create a roughly flat LSB residual background which either must be measured and removed, or else the sky image must be normalized and scaled to individual exposure background fluxes \citep{Duc2015}.  In either case, we encounter again the problem of contaminating flux from undetected or unmasked sources.  

The method's utility depends critically on the survey's observation strategy: images must be dithered to a large enough extent that the same sources do not continually fall in the same pixels on the camera from exposure to exposure (i.e., they must be larger than the LSB structures one wishes to detect), and they must be taken close enough together in time that the sky can be assumed stable across combined exposures.  In practice, the method is used to remove the time- and position-variable instrumental scattered light in addition to the true sky variation.

To test this method, we created a series of dithered exposures from our master images by taking $4096 \times 4096$~px cutouts at coordinates roughly following a $3 \times 3$ grid centred at the master images' central pixels, with a grid spacing of 2048~px (half the synthetic camera width).  We added stochasticity to this pattern by adding offsets to the central coordinates at each synthetic pointing along the $x$ and $y$ directions, randomly chosen for each direction from a uniform distribution with a maximum absolute value of 10\% the dither spacing.  We then added the same sky model to each of these synthetic exposures, altering only the noise arrays, and median combined these exposures together using five iterations of $3\sigma$ clipping.  Finally, we convolved the resulting image with a Gaussian kernel with $\sigma=273$~px (0.9\arcmin), or $1/15$ the synthetic camera size.  We chose this to mimic the smoothing scale used by \citet{Duc2015} relative to their camera's field of view.

For each of our experimental cases, we made such median-combined sky images by combining $N=3$--$15$ exposures.  Again, as a metric to assess the method's performance, for each value of $N$, we subtracted the input model sky (without noise) from the resulting average sky image and measured the median and MAD of the resulting residual images.  We show these values plotted against $N$ in Fig.~\ref{fig:dither}, with median values in the left column and MAD values in the right column.  As before, we adopted the standard error on the median as our uncertainty on the median values of the residual images, though in all experiments these uncertainties are negligible.

Each experiment once again shows some common behaviour.  As $N$ increases, both the median and MAD decrease, ultimately becoming roughly constant between $\sim10$--$15$ combined exposures \citep[similar to the behaviour described by][]{Duc2015}.  This behaviour is significantly noisier in the extended PSF case, likely due to the uneven distribution of bright stars across our master images, though the broad trends are still present.  In each case, the median flux of the residual images is always positive---because we are merely averaging exposures, the sources present in each image form a fairly smooth, LSB pedestal background \citep[as was noted by][]{ferrarese12}.  The brightness of this pedestal level depends on the kinds of sources present in the combined images: where the profiles converge, the faintest pedestal ($\sim1.5$~ADU) arises in the control case, where sources are injected without scattered light and source density is relatively low, whereas the brightest ($\sim 15$~ADU) arises in the extended PSF case, where the wings of bright stars dominate the background flux.  The cosmological background case shows a pedestal between these two, reflective of the higher source density in the images.  To account for this, the Elixer-LSB pipeline normalizes their sky images to unit flux, then scales these normalized images to individual exposure backgrounds and subtracts them.  As such, measurement of the scale factor becomes the primary source of uncertainty for this method (e.g., if the mean or median of background flux is used, it would suffer from the issues discussed in Sec.~\ref{subsubsec:meanmed}).

\begin{figure}
    \centering
    \includegraphics[scale=1.0]{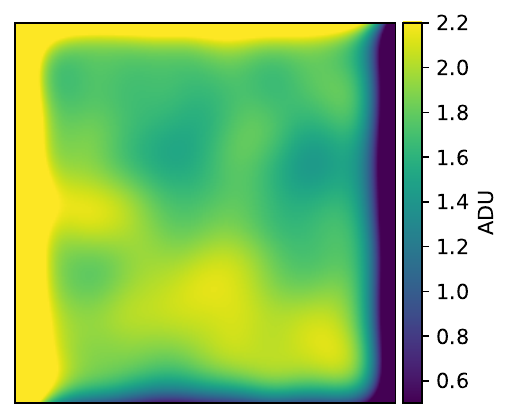}
    \caption{An example of the residuals from the median combination of dithered synthetic images.  To generate this image, we median-combined 15 such exposures, smoothed this image with a Gaussian kernel with $\sigma=273$~px, then subtracted from this image the noiseless input sky model. Artifacts along the edges result from the convolution software.  \label{fig:residual}}
\end{figure}

The values of MAD, as well, decrease roughly as $N^{-1/2}$, approaching $0$ asymptotically.  As the residual images can never be perfectly noise-free, the use of this method will also add noise to each image from which the resulting sky models are subtracted.  Smoothing helps reduce this noise on small-scales, but it will always be present at least on the smoothing scale itself.  To demonstrate this, we show an example of a residual sky image in Fig.~\ref{fig:residual}, in which peaks and troughs roughly $\sim300$~px in size are visible (alongside artifacts along the image edges resulting from the convolution software).  The inverse of such artifacts are imprinted on every image from which these sky estimates are subtracted.  The amplitude of these features is fairly small ($\sim1$~ADU, $\mu_{i} \sim 29.5$ mag arcsec$^{-2}$), but their presence could limit the accuracy of surface photometry for objects of order or larger than the smoothing kernel scale.  A well-considered dither pattern, which ensures that clusters of objects or extended objects rarely fall into the same region of the focal plane across all exposures, could help mitigate this.

In both the Elixer-LSB method and in parametric estimation, the primary source of error for sky estimation is flux contamination from non-sky emission.  The next section describes a third alternative, designed to remove this influence prior to sky estimation via difference imaging.

\subsection{Coadd-subtraction method}\label{subsec:coaddsub}

\begin{figure*}
    \centering
    \includegraphics[scale=1.0]{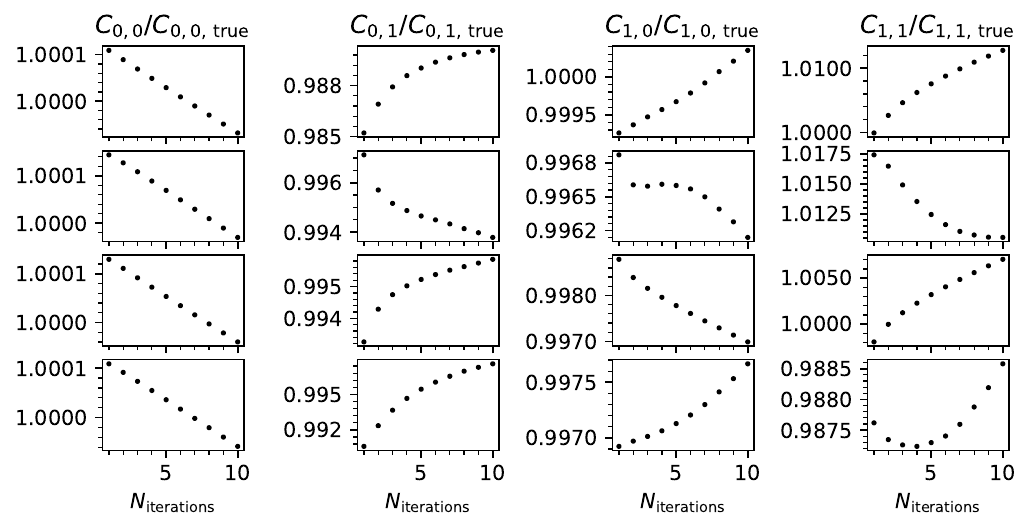}
    \caption{Evolution of four best-fit Legendre polynomial coefficients for sky estimation when iterating the coadd-subtraction procedure.  Each panel shows the best-fit Legendre coefficient (indicated at the top of each column) normalized by the true respective coefficient value for $N=10$ iterations of coadd-subtraction and re-estimation of the sky background.  Each rows shows coefficients for a different randomly selected test image.  \label{fig:evolution}}
\end{figure*}

The primary limitation of the two methods discussed in the previous sections is the tendency for flux from unmasked fixed sources to bias the sky models.  In every part of the sky, this imposes a fundamental limit on these methods which depends on the sources of such contamination---for extremely deep exposures, this will likely come from a combination of the high-redshift background (which can be corrected for empirically), faint Milky Way stars, and diffuse light reflected from Galactic cirrus (much less simple to correct for), all of which are omnipresent in the night sky, with the latter two sources becoming more prominent with approach to the Galactic plane.  Indeed, cirrus poses one of the largest problems for sky estimation; based on the all-sky dust map from Planck \citep{ade16}, a simple scaling to optical surface brightness suggests that as much as 80\% of the sky could be covered with cirrus at the $\mu_{V} \sim 30$--$31$ mag arcsec$^{-2}$ level \citep{mihos19}.  An ideal sky-subtraction therefore requires a means to decouple fixed sources from the time-variable night sky prior to estimating the latter.

This is possible using difference imaging.  From each image for which one wishes to estimate the sky, one can subtract a reference image at the same celestial coordinates, removing all fixed objects but leaving behind all time-variable flux.  This includes cosmic rays, satellite trails, some scattered light artifacts, and also the night sky emission.  This approach can take two forms: either the reference image is a single exposure from the observing run (with its own night sky emission and time-variable artifacts), or it is a coadd of all images from the observing run, either with an average sky or with preliminary models of the sky removed from each exposure prior to coaddition.  Here we investigate this last approach, as it benefits from a high signal-to-noise reference image with guaranteed matching spatial coverage for all exposures.

For this experiment, we constructed a series of 45 dithered exposures from the cosmological background master image, with a dither pattern following that described in Sec.~\ref{subsec:avskies}, but using a spacing of $1/4$ the image size (1024~px) rather than $1/2$.  We added a randomized synthetic sky to each image, using the sky parameters outlined in Sec.~\ref{subsubsec:polyfit} as a base, but altering for each exposure each coefficient value by a random number drawn from a normal distribution $\mathcal{N}(0,\, 0.05\times C_{n, m})$, where $C_{n, m}$ is the default value for coefficient $n, m$. Because we set $C_{1,1}$=0 initially, to allow for variation in this parameter, we set the default instead, arbitrarily, to $1.35 \times 10^{-6}$.

This randomized sky is not necessarily reflective of real changes to night sky emission over a set of exposures; moonlight or city lights, for example, can impose gradients which change regularly as the telescope approaches or retreats from either source, for example.  However, because we attempt to remove the skies from each image prior to coaddition, the exact manner of sky variation from image to image should not affect the results of our experiment.

We constructed a preliminary sky-subtracted coadd by masking each image to a depth of $\mu=27$ mag arcsec$^{-2}$, excluding sources with $m_{i} > 26.1$ from the masks as described in Sec.~\ref{subsubsec:meanmed}.  This deliberately ensures that the initial sky estimates were incorrect---we can only assess this method's utility by determining whether or not the sky models generated post-coadd-subtraction are closer to the true sky models than the initial estimates.  We then fit the skies from these masked images as described in Sec.~\ref{subsubsec:polyfit}, subtracted these best-fit models from each image, and registered and median-coadded the sky-subtracted images using the Astropy \citep{astropy:2018} implementation of CCDPROC \citep[Ver. 2.3.1;][]{craig22}.

The coadd, being an average of all images, has different properties (PSF, image dimensions) than the exposures used to create it.  Hence, to subtract this coadd from each image, we re-centred the coadd to match each image's central coordinates (we used only integer pixel offsets to trivialize this procedure), then cropped the coadd to the image's dimensions ($4096 \times 4096$~px).  In real images, the coadd's photometric zeropoint will also differ from those of individual exposures, which were taken under variable photometric conditions and at varying airmasses.  To simulate the impact of this on the coadd-subtraction process, we thus also scaled the cropped coadd in flux to each exposures' photometric zeropoint prior to subtraction by fitting a regression line between all image and coadd flux with values between $100 < F < 10000$~ADU above the mean sky background (which, in the coadd, is always zero).  For our images, this scale factor was always nearly one ($\sim0.997$, on average), as we created each synthetic image with the same zeropoint $ZP=33.1$ (Sec.~\ref{sec:synthetic}), demonstrating that this scaling method worked as expected.  We finally multiplied each cropped coadd by this scale factor and subtracted it from its associated image.

\begin{figure}
    \centering
    \includegraphics[scale=1.0]{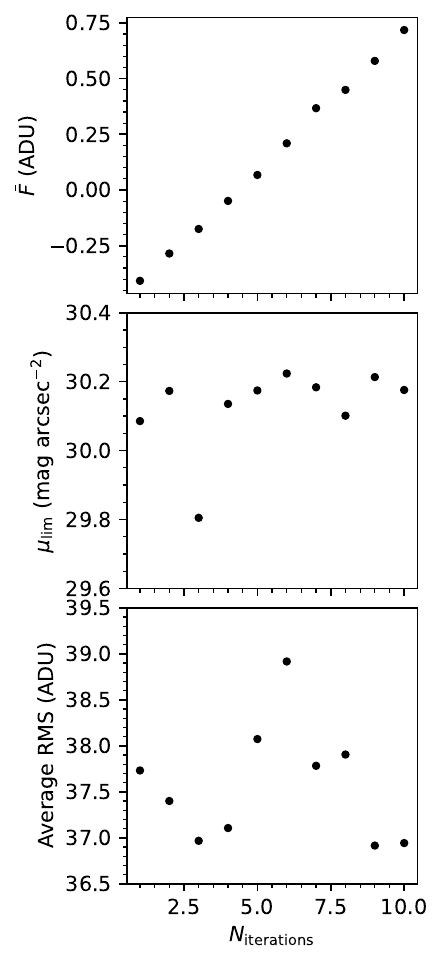}
    \caption{Evolution of the background mean and noise in sky-subtracted coadds while iterating the coadd-subtraction procedure.  From top to bottom, we show the mean background flux, the limiting surface brightness ($1\sigma$, on 10\arcsec$\times$10\arcsec \ scales), and the average pixel-to-pixel root-mean-square as a function of the number of sky re-estimation iterations.  \label{fig:coadd_noise}}
\end{figure}

Not all synthetic sources cleanly subtract out with the subtraction of the coadd.  This is due to slight differences between the coadd PSF and that of the individual exposures.  To re-estimate the skies, we thus masked our coadd-subtracted images to a much shallower depth of $\mu=24$ mag arcsec$^{-2}$ to ensure that these leftover artifacts did not contaminate the models, then fit the unmasked pixels in each difference image using the method described in Sec.~\ref{subsubsec:polyfit}.  Finally, we subtracted these new sky models from each image, and we reconstructed the coadd using these newly sky-subtracted images.

If the method is viable, iteration of this procedure should result in the best-fit sky coefficients converging toward their input ground-truth values.  We therefore repeated the process described above $N=10$ times, recording for each synthetic exposure the best-fit sky coefficients at each iteration.  We show the progression of these coefficients for a random sample of four images in Fig.~\ref{fig:evolution}.  Each column of this figure shows the progression of a different coefficient, with coefficient values per iteration normalized by the true coefficient value.  Each row shows this progression for a different image.  Convergence is reached when each panel's evolutionary curve reaches unity.

Fig.~\ref{fig:evolution} clearly demonstrates that the best-fit coefficients continually drift with each iteration.  This suggests that using a sky-subtracted image coadd as a reference is not appropriate: evidently, any artifacts present in the initial coadd from errors in the initial sky-subtraction propagate into each subsequent sky estimation, resulting in continually drifting sky models.  We do not demonstrate it here, but we found the same trend when repeating this method on real images from the Nordic Optical Telescope \citep[observed as part of the study by][]{rautio22} using second-order polynomial fits to the image backgrounds.  \citet{roman23} attempted a similar approach when initially reducing their data, again with poor results.

Additionally, we investigated the evolution of the coadd backgrounds as a function of iteration number.  An ideal coadd background should have a mean flux of zero, with noise characteristics dependent on the number of combined exposures in each part of the coadd.  To measure the background properties, we first masked each coadd to a depth of $\mu=31$ mag arcsec$^{-2}$ using a coadd created by combining dithered images with models only (no noise or skies).  We then measured each coadd's average background flux in ADU (measured as the median of the median fluxes within $N=1000$ 10\arcsec$\times$10\arcsec \ boxes distributed across the coadd, ignoring masked pixels), its limiting surface brightness (the standard deviation of those median fluxes), and its mean pixel-to-pixel root-mean-square (RMS, the mean standard deviation of the flux values within individual boxes).

We show the evolution of mean background flux, limiting surface brightness, and average RMS as a function of iteration number in the top, middle, and bottom panels of Fig.~\ref{fig:coadd_noise}, respectively.  While we find no clear trend in either limiting surface brightness or RMS, the average flux in the coadd backgrounds steadily increases with each iteration, again in a roughly linear fashion.

The reason for this behaviour is linked to the behaviour demonstrated in Sections \ref{subsubsec:meanmed} and \ref{subsubsec:polyfit}.  If the error in sky estimation was random (including over- and under-estimation), the central limit theorem would suggest that with enough coadded exposures, the background in the coadd would be approximately zero with no significant structure.  However, when using traditional sky estimation techniques, the error is instead heavily skewed toward over-subtraction, leading to a coadd with an over-subtracted background.  This accounts for the near linear drift in sky model brightness with each iteration, whose direction depends on the type of clipping algorithm used in the coaddition and the severity of the over-subtraction for each exposure.

Correcting for EBL and scattered light could reduce this drift in real images, but would not completely resolve the problem.  We verified this by repeating the procedure described in this section using the control master image (hence excluding entirely any EBL and scattered light contribution), wherein we found the same kind of fit parameter drift demonstrated in Fig.~\ref{fig:evolution}, but purely contributed by flux leakage through masks.  Any additive source would yield this behavior, including less easily characterizable sources such as reflections and Galactic cirrus.  Therefore, we must conclude that while it seems promising at first glance, the use of a sky-subtracted coadd for isolating night skies from images does not improve sky estimation over a simple masking and fitting procedure.


\section{Discussion}\label{sec:discussion}

\begin{figure*}
    \centering
    \includegraphics[scale=1.0]{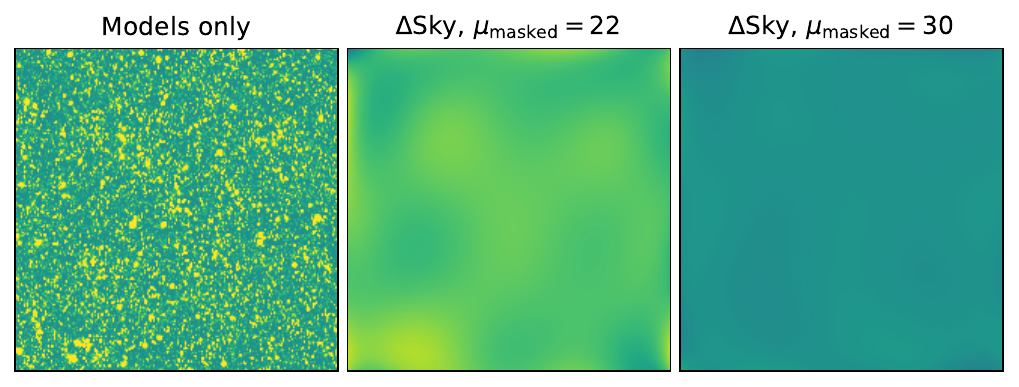}
    \caption{Demonstrating the tendency for complex sky models to over-subtract sky in the vicinity of highly clustered (on the sky) bright sources.  The left panel shows the test image, with models only (no sky or noise).  The centre and right panels show the difference between the true injected sky (first order polynomial) and a sixth-order polynomial fit to the injected sky with all sources masked to 22 mag arcsec$^{-2}$ and 30 mag arcsec$^{-2}$, respectively.  All images are scaled between -10 and 10 ADU.  \label{fig:overfit}}
\end{figure*}

\subsection{Masking and parametric sky estimation}\label{subsec:maskconcl}

Our experiments have quantified the sensitivity of sky models to contamination from fixed astrophysical (or other) sources.  Any unmasked non-sky flux in an image has a tendency to brighten measured sky models, with the level of bias depending on the amount, intensity, and distribution of the unmasked flux.  The strongest such bias in our experiments comes from scattered light around foreground stars: the extended stellar PSF, in our synthetic images, creates an LSB background with surface brightnesses 28--29 mag arcsec$^{-2}$ even in the fairly low-density regions of the sky on which we based our synthetic catalogue.  This surface brightness is not fixed, of course, but is PSF-dependent.  Regardless, if this flux is not removed prior to sky estimation, the color variation in stars can result in artificial color gradients appearing in the backgrounds of sky-subtracted images which can influence the photometry of diffuse flux with large angular size \citep[see, e.g.,][]{roman20}.

Removal of this stellar scattered light is thus paramount for any deep, LSB-targeted surveys, and is beneficial even for surveys targeting point sources or higher surface brightness objects.  Scattered light from the comparatively lower surface brightness and often fainter galaxy population should not pose as much of a problem (Sec.~\ref{subsubsec:meanmed}, \ref{subsubsec:polyfit}, and Fig.~\ref{fig:meanmedian}, \ref{fig:planefit}), meaning that if the PSF wings are removed from only stars, masking and modelling skies could remain a viable approach even for surveys with depths comparable to that expected of LSST (29--30 mag arcsec$^{-2}$, $3\sigma$ on 10\arcsec$\times$10\arcsec \ scales).  However, this approach will still fail in regions with high source-density; the fewer the pixels available for sky estimation, the deeper the masks must reach in order to recover the sky to the same accuracy (Fig.~\ref{fig:convergence}).  Some optical systems show strong scattered light or noticeable reflections originating from extragalactic objects as well \citep[see, e.g., Fig.~6 of][]{Duc2015}.  Masking and parameterized estimation thus has a fundamental limit, which may be rapidly approaching as surveys probe to deeper and deeper magnitudes.

The tendency toward sky over-estimation applies not only to entire images, but can apply locally as well.  For example, many sky-subtraction algorithms estimate local means in regular bins, then interpolate or fit these binned means to generate the full image sky models \citep[e.g.,][]{aihara18b, morganson18}.  If some of those local means are biased by unmasked flux, this can result in local over-subtraction.  This can be beneficial when LSB flux from extended objects biases the photometry of the survey's target objects, but is evidently undesirable for LSB-targeted surveys.  We demonstrate this in Fig.~\ref{fig:overfit}.  Here, we generated a synthetic exposure from our cosmological background experiment, injecting a model sky into it with the same parameters described in Sec.~\ref{subsubsec:polyfit}.  We then masked all sources in the image to depths of 22 and 30 mag arcsec$^{-2}$ and over-fit the unmasked sky pixels using a sixth-order Legendre polynomial.  The centre and right panels in the figure show the difference between the over-fit sky models and the true sky models for the two mask depths.  We show source models in the left panel, without sky or noise to more clearly demonstrate the source distribution.  Each panel is scaled to the same flux limits (-10 to 10 ADU) to facilitate comparison.  When masks are deep, even an over-fit sky closely resembles the true sky, but for shallow masks the residuals clearly follow the distribution of models, showing peaks near clusters (on the sky) of model galaxies and troughs in regions of comparatively low model density.

This behaviour explains the local over-subtraction produced by the HSC Public Data Release 1 pipeline \citep{aihara18b, aihara19}, where sky was estimated using a sixth-order Chebyshev polynomial.  We expect it will occur whenever small bin sizes are used in conjunction with complex sky models \citep[including, e.g., interpolation algorithms such as that used by Source Extractor;][]{bertin96, bertin10b} unless fixed sources are masked to extremely deep levels.  Even then, the risk of over-subtracting any unmasked LSB flux remains.  Peaks local to clustered sources still arise when masking to 30 mag arcsec$^{-2}$, with amplitudes of $\sim0.4$~ADU ($\mu_{i} \sim 30.6$ mag arcsec$^{-2}$).  The nature of LSB flux is such that even a tiny amount of linear over-subtraction can result in noticeable over-subtraction in surface brightness units: for example, the difference between 29 and 29.5 mag arcsec$^{-2}$ in our synthetic images, scaled using HSC photometric zeropoints, is only $\sim0.5$~ADU.

Generally, pipelines employing a fitting procedure should strive to use as simple a sky model as feasible while masking sources to as deep a level as possible---which will involve a compromise with the number of sky pixels available for modelling---else sky over-estimation will preferentially remove flux from both isolated extended objects and from clusters of more compact objects.  How this is achieved is not trivial in real images, given that the depth of one's detections (hence, masking) is contingent on accurate removal of sky flux in the first place.  Iteration of sky estimation and masking may be necessary to converge to the most accurate possible sky model for any given image.

\subsection{Non-parametric estimation}\label{subsec:ditherconcl}

Non-parametric sky estimation provides an alternative to masking and modelling.  Through our experimentation, we found that the method employed by the Elixer-LSB pipeline \citep{ferrarese12, Duc2015}---median-combining dithered exposures and smoothing the resulting combined image with a large kernel---is an effective and fast non-parametric alternative, though it too suffers effects from contaminating flux.  Sources in the images combined to create the sky models tend to create a roughly flat LSB residual background, which either must be measured and removed, or else the sky image must be normalized and scaled to individual exposure background fluxes \citep{Duc2015}.  In either case, we encounter again the problem of contaminating flux from undetected or unmasked sources.

Another concern with this particular method is the addition of noise: as sky models are generated by combining and smoothing exposures, each sky model so produced contains low-level artifacts at the smoothing scale, which add noise to each exposure from which they are subtracted.  This imposes a noise limit on the binning scale (on top of that imposed by the usual Poissonian background, read noise, and other common sources), which may not be ideal if one's astrophysical targets of interest are of similar scale or larger.  This can be mitigated somewhat by combining more images or by using larger smoothing kernels, but the former hinges on the survey cadence (which must be adapted specifically for the method), and too large a kernel risks over-smoothing the sky and washing out small-scale features, resulting in a less locally accurate sky model and thus an increase in correlated noise (Fig.~\ref{fig:polyres}).

All of this assumes, as well, that the sky is indeed almost constant across all combined exposures.  For large-scale, multi-purpose surveys like LSST, whose cadence depends primarily on its key science goals \citep{bianco22}, this method will likely be impractical due to too large a variability in the sky emission from image to image for a given region\footnote{The baseline cadence recommendation optimises for sky coverage per night, leading to few exposures taken nightly in the same area of the sky: https://pstn-055.lsst.io/}.  Space-based surveys such as Euclid may be able to employ this method, as the foreground "sky" emission present in space-based imaging is dominated by diffuse interplanetary dust (zodiacal light), which is smooth in appearance on large angular scales and varies only seasonally due to the Earth's orbit about the sun \citep[altering the line of sight through the dust;][]{kelsall98}.  For the Euclid Wide Survey, each field will be observed sequentially, not seasonally, with four exposures taken at four dither pointings per field \citep{scaramella22}, making this a potentially viable approach.  For a more thorough discussion of sky-subtraction in space-based images, see Sec.~2.6 of \citet{borlaff19}.

\subsection{Isolating the night sky from fixed sources}\label{subsec:coaddconc}

The ideal solution would be to remove time-invariant sources first, before any attempt is made at sky estimation.  However, in practice, this proposition is circular: without removing the night sky emission, faint objects cannot be detected, but without removing faint objects, one's estimate of the night sky emission will be biased.  This can be circumvented somewhat using a time-consuming iterative approach to sky estimation, in which masks are improved post-sky-subtraction, and the sky is re-estimated using the improved masks \citep[e.g.,][]{watkins16, mihos17}.  

We investigated a possible way to circumvent this through the alignment and subtraction of a preliminary sky-subtracted image coadd.  However, we found that because errors in the initial sky estimations are not random, instead always skewing toward over-subtraction, those errors propagate continuously into any subsequent sky estimations made from coadd-subtracted images (Fig.~\ref{fig:evolution}, \ref{fig:coadd_noise}), rendering such a procedure ineffective on top of a more traditional sky estimation.  In real images, some of this drift can, of course, be accounted for by including a flux offset to the sky models based on background galaxy counts \citep[e.g.,][]{ji18}, among other corrections (e.g., PSF subtraction).  However, other less uniform and less easily parametrisable sources of contamination (moonlight, faint reflections or flares, Galactic cirrus, etc.; see the following section) will also tend push sky models brighter.  These sources' more complex structure would lead to more complex inaccuracies in the models, which will again be imprinted on all exposures once the coadd is subtracted \citep[e.g., see the discussion of extremely low-level scattered light from Mars in][]{watkins14}.  If the method only works when the initial sky estimate is perfect, the method becomes redundant.  Previous efforts at similar methods have reached similar conclusions \citep[e.g.,][]{kluge20}.

One potential alternative is removal of contamination through source modelling.  \citet{kelvin23} discusses this approach; in their experiments, they ran a detection algorithm on their synthetic images, fit S\'{e}rsic models to each detected source brighter than some relatively bright threshold value, then subtracted the best-fit models of each source from the images, effectively modelling out the influence of diffuse flux from the injected sources.  This proved quite effective in reducing the brightness of the modelled skies (see, e.g., their Fig.~3), serving as a successful proof of concept.  However, real sources are more complex than single S\'{e}rsic profiles (particularly at surface brightnesses at which the stellar halo dominates), so it remains to be seen how well this approach would work on real images.  It also would not be an effective approach to handle the impact of Galactic cirrus, which is not easily modelled but is present at some level in all astronomical images \citep[e.g.,][]{mihos19}.

Another alternative would be similar to the coadd-subtraction approach we outline in Sec.~\ref{subsec:coaddsub}, but avoiding any preliminary sky estimation.  This can be done in the following manner (Al Sayyad, priv. comm.).  For a given part of the sky, one selects as a reference image a single exposure with similar coordinates.  Every other exposure containing that same part of the sky is then subtracted from this reference image: this yields images with all fixed sources removed (where the two images overlap), as well as backgrounds composed of the difference between the reference image background and the image backgrounds (hereafter, $\Delta$sky).  If one then models this $\Delta$sky image (extrapolating as needed to where the images do not overlap, and accounting for, e.g., CCD-to-CCD amplifier offsets or other such artifacts) and adds this model to each image, it generates images with backgrounds effectively matched to that of the reference image.  Registering and coadding all background-matched images, assuming the matched sky models are properly aligned in celestial coordinates, would yield a coadd with a background matched to that of the reference frame (expanded to the larger angular scale covered by the coadd).  This can then be estimated and removed at the coadd level, where sources are more easily detectable and masked given the coadd's higher signal-to-noise ratio.  Explicitly testing this method is beyond the scope of this paper; however, we can speculate that, while this approach would still be limited by all of the factors discussed throughout this paper (e.g., flux contamination from undetected sources, cirrus, or mask leakage), it may be more efficient than other masking and model-based approaches, as masking would only be required on the coadd and not on individual exposures.

\subsection{Other sources of contamination in sky models}\label{subsec:other}

As we reiterated throughout this paper, we conducted our experiments using highly idealized synthetic images, to facilitate interpretation and to isolate the mechanisms affecting the sky estimation algorithms we tested.  Our images contain only Gaussian read noise and shot noise from the input sky models, as well as stationary astronomical sources (stars and galaxies), with time-invariant luminosities and identical PSFs from experiment to experiment, without including realistic exposure time-induced shot-noise for each source.  Presence of correlated noise, including shot noise from unresolved, barely detected faint sources, can impact mask generation, where true noise peaks are indistinguishable from faint astrophysical flux.  The masking software itself can impact the sky solution depending on how it treats image noise \citep[e.g.,][]{kelvin23}.

We also assumed that every pixel in our images had the same gain (i.e., the implied flat-field correction was perfect), with no oddities such as filter vignetting or amplifier-induced directional flux rolloffs.  However, we assume that such artifacts need not be taken into account here, as they are best treated separately to sky estimation, using specific calibration images such as bias frames and dome flats.  Some calibration artifacts are sensitive to sky emission, however; in large, multi-detector systems like HSC or LSST, sensitivity in certain wavelengths can noticeably vary from CCD to CCD, and variation in filter transparency across the focal plane can impose patterns in image backgrounds with amplitudes dependent on the illuminating source \citep[e.g., the HSC sky frames;][]{aihara19}.  \citet{aihara19} measured these sky frames for HSC via combination of a large number of dithered exposures, similar to the Elixer-LSB method discussed in Sec.~\ref{subsec:avskies}, thus suggesting it may suffer similar problems of noise and scaling.  However, the structure of these patterns is stable on much longer timescales than the structure of the night sky emission, and is driven by the optical components (meaning it does not vary with telescope pointing position), allowing for combination of a much larger number of exposures, reducing noise to manageable levels.  This leaves only the problem of deriving an appropriate scale factor for each exposure.  Fringe patterns behave in a similar way.

Finally, there exist additive, scattered light artifacts like ghosts, glints, flares, reflections, satellite or asteroid trails, and other ephemera like cosmic ray trails, dead pixels, and so on.  Even assuming calibration and instrumental signature removal is perfect, all of these additive artifacts can impact one's sky estimation as well by injecting additional flux into images at varying surface brightnesses; extrapolating from our results, such a flux injection will tend to bias sky models brighter, unless the sources themselves are either masked (limiting still further the pixels available for accurate sky estimation) or modeled and removed.

Reflections in particular are problematic, as they are part of the PSF insofar as they are associated with every point source and have fixed sizes which depend on the light path through the telescope components \citep{slater09}.  They are most noticeable around naked-eye stars, but still occur at lower surface brightness around all astrophysical sources, injecting another LSB pedestal into images which may not be readily detectable.  Ideally such artifacts are mitigated via anti-reflection coating, or are removed at the source via modelling and subtraction prior to sky estimation.  This is in some ways simpler than modelling and removing galaxies, as PSFs and reflections have fixed, derivable properties \citep[e.g.,][]{slater09}, though for complex optical systems like HSC or LSST this derivation is not trivial.

Artifacts from more isolated events, such as glints or satellite trails, are much simpler to remove; here, difference imaging (such as the methods reviewed in the previous section) can be beneficial, allowing for more accurate modelling of such artifacts via removal of fixed astronomical sources.  Therefore, while additive artifacts can bias sky estimates in a manner similar to that of fixed astronomical sources, accounting for them is in some ways more feasible than accounting for, e.g., faint Milky Way stars or diffuse Galactic cirrus \citep[altough for an innovative means to include cirrus in sky models, see][]{liu23}, whose measurability relies on accurate night sky removal in the first place.


\section{Summary and Recommendations}\label{sec:conclusions}

Using idealized synthetic images, we tested the limits of two sky-subtraction methods commonly employed for low-surface-brightness imaging surveys: masking of fixed astronomical sources and parametric sky estimation, and median-combining and smoothing of dithered exposures closely spaced in celestial coordinates and time.  We tested each using four experiment types:

\begin{itemize}
    \item Control: images populated with model HST-COSMOS survey galaxies and HST-COSMOS and SDSS stars convolved with a Moffat profile, to establish a low-density, low-scattered-light baseline
    \item Extended PSF: the same sources as Control but convolved with the 7\arcmin \ (8\arcmin, using our pixel scale) HSC PSF from \citet{Montes2021}, to study the effects of scattered light
    \item Star-subtracted: the same sources as Control, where galaxies are convolved with an 8\arcmin \ PSF but stars are convolved with a Moffat profile, to simulate an idealized post-star-subtracted image such as those employed in LSB-oriented data reduction pipelines
    \item Cosmological background: as star-subtracted, but including also a high-redshift, faint source background using a light cone from the Horizon-AGN simulation, to study the effects of undetected faint sources
\end{itemize}

For up-coming deep-wide surveys such as LSST, we find that traditional sky-subtraction methods like masking and parametric modelling remain efficient and effective methods for removing night-sky emission without affecting LSB flux, at least in fields without significant contamination from extended LSB sources like Galactic cirrus (a regime not probed in this study).  One must take care to mask fixed sources to deep enough levels (preferably below 31 mag arcsec$^{-2}$ for surveys with LSST-like expected depths, to keep systematic sky over-estimation to levels below 30 mag arcsec$^{-2}$; Fig.~\ref{fig:meanmedian} and \ref{fig:planefit}) and to avoid over-fitting unmasked pixels using complex sky models (Fig.~\ref{fig:overfit}).  In crowded fields, sky brightness will likely still be noticeably over-estimated, but so long as a simple model is employed, this over-estimation will be a global one across the focal plane rather than localized to extended sources or clustered sources, where its impact on photometry would be much more complex and harder to correct for.

Non-parametric methods, such as the median-combining of dithered exposures \citep{tyson90, ferrarese12, Duc2015}, can also be effective for surveys with depths in excess of 29 mag arcsec$^{-2}$, so long as the survey cadence allows for the combination of a large enough number \citep[10--15 ideally, although $\sim7$ is a viable compromise if few such exposures are available;][]{Duc2015} of exposures of a single part of the sky over a short enough period of time, such that the sky can be assumed stable across those exposures.  As it requires no masking, the method is fast, and it accounts for detection of very extended LSB structures, so long as the dithering offsets are large enough.  The primary concern with this method aside from the survey cadence may be the addition of noise on the scale of the smoothing kernel used to reduce the pixel-to-pixel noise in the combined image (Fig.~\ref{fig:residual}), but photometry of sources with sizes smaller than this kernel should be unaffected.

The utility of each of these methods relies on the prior subtraction of the extended wings of the PSF around bright stars and the removal of the pedestal imposed by extragalactic background light.  Scattered light from stars specifically produces a bright (in our synthetic images, 28--29 mag arcsec$^{-2}$, even in relatively low-density fields; top-right panel of Fig.~\ref{fig:experiments}), complex background on top of the night-sky emission.  Attempting to remove this background alongside the night-sky emission necessitates complex, local sky models, and so risks over-subtraction of the flux of extended or clustered sources, as described above.  PSF-subtraction is a necessary early step for any LSB-oriented data reduction pipeline.  Indeed, PSF deconvolution of all detected objects, on top of an EBL correction, would be ideal.

We find that the only means to perfectly estimate night sky flux is via the prior subtraction of all static objects (stars, galaxies, Galactic cirrus, and so on).  Our attempts at doing this via difference imaging with a preliminary sky-subtracted coadd failed, as any errors made during the preliminary sky subtraction used to generate the coadd are imprinted into the coadd background, which are then imprinted in reverse into the images upon coadd-subtraction.  The only way to avoid this is to subtract the correct sky models from all exposures in the first place, which negates the need to re-estimate the skies after coadd-subtraction. A background-matching technique (Sec.~\ref{subsec:other}) could serve as a useful middle ground between the ideal night-sky subtraction we envisioned when developing our coadd-subtraction method and the more traditional techniques.  This could cut processing time by allowing one to model the sky by masking only the coadd, but the exact pitfalls and benefits of this particular method for now will remain a subject for future research.

Broadly, while pursuit of a perfect sky-subtraction algorithm remains a future endeavor, existing algorithms can still model night sky emission accurately enough for up-coming deep, wide surveys, without blending it with fixed astrophysical flux, when the overall observation strategy and data reduction pipeline is built with LSB science in mind.  Surveys (including deep drilling fields of up-coming surveys like LSST) probing the $\mu > 31$ mag arcsec$^{-2}$ regime may require more innovative solutions, however, to fully exploit the unprecedented depth.


\section*{Acknowledgements}
We thank the anonymous referee for their detailed and well-considered evaluation of our work, which greatly improved the manuscript.  We thank both the members of the LSST Low-Surface-Brightness Working Group, particularly those members participating in the Low Surface Brightness Challenge 1 group, and the members of LSST:UK for providing valuable feedback and ideas throughout the course of this work. We thank Yusra Al Sayyad, Robert Lupton, Dan Tanaru, and Sophie Reed from the Princeton Rubin Data Management team for sharing their time, ideas, and expertise with us during the course of this work. AEW and JCM also thank Paul Harding for exhaustive discussions about optimal methods for sky subtraction.  AEW and SK acknowledge support from the STFC [grant numbers ST/S00615X/1 and ST/X001318/1]. AEW acknowledges travel support provided by STFC for UK participation in LSST through grant ST/S006206/1. SK acknowledges a Senior Research Fellowship from Worcester College Oxford. JR acknowledges funding from University of La Laguna through the Margarita Salas Program from the Spanish Ministry of Universities ref. UNI/551/2021-May 26, and under the EU Next Generation.

This research has used the DiRAC facility, jointly funded by the STFC and the Large Facilities Capital Fund of BIS, and has been partially supported by grant Spin(e) ANR-13-BS05-0005 of the French ANR. This work was granted access to the HPC resources of CINES under the allocations 2013047012, 2014047012 and 2015047012 made by GENCI. This work is part of the Horizon-UK project. It has also made use of the Horizon cluster on which the simulation was post-processed, hosted by the Institut d'Astrophysique de Paris. We warmly thank S.~Rouberol for running it smoothly.

This paper makes use of software developed for the Vera C. Rubin Observatory.  We thank the Rubin Observatory for making their code available as free software at \url{http://pipelines.lsst.io/}.

This paper makes use of data collected at the Subaru Telescope and retrieved from the HSC data archive system, which is operated by the Subaru Telescope and Astronomy Data Center (ADC) at NAOJ.  We are honored and grateful for the opportunity of observing the Universe from Maunakea, which has cultural, historical and natural significance in Hawai'i.

This work made use of Astropy:\footnote{http://www.astropy.org} a community-developed core Python package and an ecosystem of tools and resources for astronomy \citep{astropy:2013, astropy:2018, astropy22}, as well SciPy \citep{scipy20}, NumPy \citep{harris20}, and Matplotlib \citep[Ver. 3.7.1;][]{hunter07, caswell23}.

For the purpose of open access, the authors have applied a Creative Commons Attribution (CC BY) licence to any Author Accepted Manuscript version arising from this submission. 


\section*{Data Availability}

The software we built to create our synthetic images is available at the LSST:UK Github repository\footnote{\url{https://github.com/lsst-uk/sky-estimation-WP3.7/tree/master/syntheticImages}}.  Our master images, as well as the source parameter tables used to generate them, are hosted on Zenodo\footnote{\url{https://doi.org/10.5281/zenodo.8192052}}, with accompanying documentation.  We selected our synthetic source parameters using the publicly available \GalSim COSMOS training sample\footnote{\url{https://galsim-developers.github.io/GalSim/_build/html/real_gal.html}}, the COSMOS2015 source catalogue \citep{laigle16}, and the SDSS Data Release 12 \citep{alam15}.  The extended HSC PSF derived by \citet{Montes2021} is available on Dropbox\footnote{\url{https://www.dropbox.com/sh/rvscig6ucaa0bx5/AACmzJ1ANZ6iOdEOTGEAT1_8a?dl=0}} or upon request to those authors, in the event the directory's contents alter in the future.


\bibliographystyle{mnras}
\bibliography{references} 


\appendix


\bsp
\label{lastpage}
\end{document}